\documentclass[%
reprint,
amsmath,amssymb,
aps,
prb,
floatfix,
]{revtex4-1}

\usepackage{graphicx}
\usepackage{dcolumn}
\usepackage{bm}
\usepackage{amsmath}
\usepackage{subfigure}
\usepackage{color}

\begin{document}

\preprint{UI-NLO/0002}

\title{Collective Modes of Massive Dirac Fermions in Armchair Graphene Nanoribbons}


\author{David R. Andersen}
 \email{k0rx@uiowa.edu}
 \altaffiliation[Also at ]{Department of Physics and Astronomy, The University of Iowa.}
\author{Hassan Raza}%
 \email{hassan-raza@uiowa.edu}
\affiliation{%
Department of Electrical and Computer Engineering\\ The University of Iowa, Iowa City, IA 52242, USA
}%

\date{\today}

\begin{abstract}
We report the plasmon dispersion characteristics of intrinsic and extrinsic armchair graphene nanoribbons of atomic width $N = 5$
using a $p_z$-orbital tight binding model with third-nearest-neighbor (3nn) coupling. The coupling parameters are obtained by fitting the 3nn
dispersions to that of an extended H\"uckel theory. The resultant massive Dirac Fermion system has a band gap $E_g \approx 64\ meV$.
The extrinsic plasmon dispersion relation
is found to approach a common dispersion curve as the chemical potential $\mu$ increases, whereas the
intrinsic plasmon dispersion relation is found to have both energy and momentum thresholds.
We also report an analytical model for the extrinsic plasmon
group velocity in the $q \rightarrow 0$ limit.
\end{abstract}

\pacs{73.20.Mf}
\maketitle


\section{\label{sec:background}Introduction}

\noindent Graphene exhibits massless Dirac Fermions \cite{Wallace,Geim, Neto08, Raza_book} with semi-metallic behavior, for which the collective carrier modes in the form of plasmons have been a topic of study \cite{Andersen, Sarma, Rana, Gangadharaiah, Jablan, Mishchenko}. When graphene is patterned in the form of an armchair graphene nanoribbon (acGNR) \cite{Nakada96, Brey06, Son, Raza08, Raza08_ac_prb}, two important deviations occur. First, the acGNRs develop a band gap and second the dispersions do not remain linear anymore and hence the electrons and holes behave as massive Dirac Fermions.  

While acGNRs of atomic widths $N$ of $mod(N,3)=0,1$ exhibit significant band gap opening irrespective of theoretical model, acGNRs with $mod(N,3)=-1$ have zero band gap and massless dispersion within the continuum and the first nearest-neighbor $p_z$-orbital tight binding (1nn pzTB) model \cite{Nakada96, Brey06, Saito, dra11a}. One has to use more detailed methods like density functional theory (DFT), \cite{Son} extended H\"uckel theory (EHT) \cite{Raza08, Raza08_ac_prb} or beyond 1nn TB model to get a more detailed band structure. Although there are quantitative differences amongst these methods, nonetheless qualitatively these methods converge upon massive Dirac Fermions with a band gap opening for $mod(N,3)=-1$ acGNRs. The band gaps predicted by EHT for these acGNRs are of the order of a few tens of meV for extremely narrow ribbons, and decreases as the width of the nanoribbon increases \cite{Raza08_ac_prb}.

\begin{figure}[htb]
\centerline{\includegraphics[width=8.5cm]{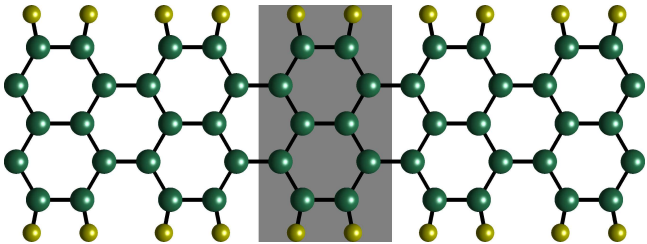}}
\caption{(Color online) Ball and stick model of the hydrogenated acGNR$5$. x- (y-) is the longitudinal (transverse) direction. Atomic visualization is done by H\"uckel-NV \cite{HuckelNV}.}
\label{fig:acGNR5geometry}
\end{figure} 

In this paper, we examine the plasmon dispersion in intrinsic and extrinsic acGNR with atomic width $N = 5$ (acGNR5) by using a third-nearest-neighbor (3nn) pzTB model, benchmarked with EHT, within the random phase approximation (RPA) as discussed in Sec. II. We discuss the plasmon dispersion results in Sec. III followed by the conclusions.  

\section{\label{sec:model}Theoretical Model}

The unit cell for a hydrogen passivated acGNR$5$ is highlighted in Fig. \ref{fig:acGNR5geometry} that contains 10 carbon and 4 hydrogen atoms. The unit vector is given as $\vec{a} = d \widehat{x} = 3 a_{cc} \widehat{x}$, where $a_{cc} = 1.42 \mathrm{\AA}$ is the carbon bond length. The pzTB Hamiltonian of the unit cell is a $10 \times 10$ matrix containing 3nn couplings. We transform the real-space Hamiltonian to the reciprocal space $H(k)$ to calculate
the eigenvalues $E_i(k)$ and eigenfunctions $c_i^{(\alpha)} (k)$ for the eigenstate $i=1,2,\ldots,10$, where $i$ is the band index and $\alpha$ represents the atomic location. The band index ranges from $i=1 \; (i=\overline{1})$ corresponding to the lowest-lying conduction (highest-lying valence) band and $i=5 \; (i=\overline{5})$ corresponding to the highest-lying conduction (lowest-lying valence) band.
One finds that the electron-hole symmetry is broken due to finite 2nn and 3nn couplings. 

\begin{figure}[htb]
\centerline{\includegraphics[width=8.5cm]{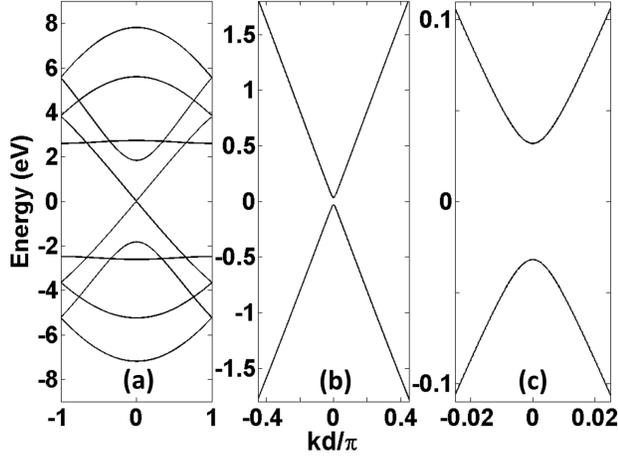}}
\caption{Carrier dispersion for acGNR5 calculated using the 3nn tight binding Hamiltonian with hopping parameters given in Table \ref{tab:tbparams}.
Panel (a) shows the complete 10-band structure, and panels (b) and (c) show progressively more detail in the dispersion near the gap of $E_g \approx 64$ meV over two different ranges of momentum.
In addition to the gap, asymmetry between the conduction and valence bands has been introduced by the non-zero 2nn and 3nn coupling parameters.}
\label{fig:bandstructure}
\end{figure}

The band structure for acGNR$5$ is shown in Fig. \ref{fig:bandstructure}. 3nn tight-binding parameters ($E_0$,$t_0$,$t_1$,$t_2$) for the fit of the acGNR5 nanoribbon to the EHT \cite{Raza08_ac_prb} are reported in Table \ref{tab:tbparams}. Parameters are obtained by fitting the top three valence bands and bottom three conduction bands of the EHT data to the 3nn pzTB band structure at 51 k-points uniformly spaced across the Brillioun zone. Fitting is accomplished using a least-squares algorithm and no geometric relaxation of the bond lengths is incorporated. This set of hopping parameters agrees well with Ref. \cite{wohlthat} for the $t_0$ and $t_1$ parameters.  However, the 3nn hopping parameter ($t_2$) we report is significantly smaller, due to the smaller gap predicted by EHT \cite{Raza08, Raza08_ac_prb} when compared with DFT results \cite{Son}.

The band structure computed using the 3nn Hamiltonian with hopping parameters from Table \ref{tab:tbparams} is shown in Fig. \ref{fig:bandstructure}.  
The carrier dispersion characteristics of the $i=\overline{1}, \, 1$ (valence, conduction) bands in this system show a finite gap of $E_g \approx 64$ meV and represent a massive Dirac Fermion system with a dispersion
characterized by the relation:
\begin{equation}
E_{ki} = \pm \sqrt{(m_{0} v_{Fi}^2)^2 + (\hbar v_{Fi} k)^2}
\label{eq:diracfermion}
\end{equation}
where $v_{Fi}$ is the Fermi velocity for the $i=\overline{1}, \, 1$ (valence, conduction) bands, and the $+ \, (-)$ sign is chosen for the conduction (valence) band. The band gap $E_g$ corresponds to a relativistic rest mass of the massive Dirac Fermion system of $m_0=E_g/2 v_{Fi}^2$.

\begin{figure}[htb]
\centerline{\includegraphics[width=8.5cm]{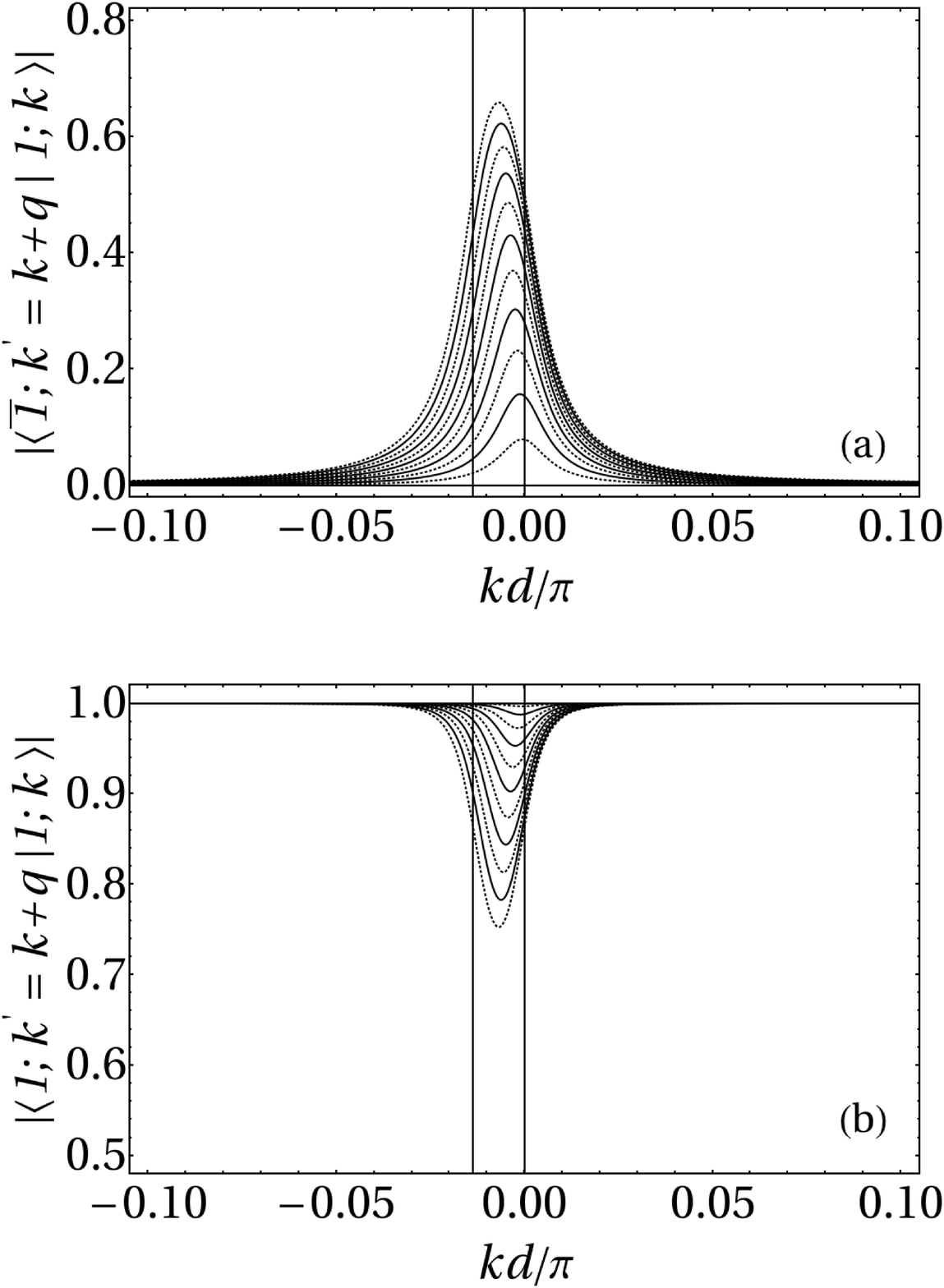}}
\caption{3nn tight-binding overlap integral computed for (a) the interband transition between conduction and valence bands,
and for (b) the intraband transition of the conduction band along the line $k^\prime = k+q$.
In each panel, a family of 12 curves is shown for $0 \le q \le q_{max}$ where $q_{max}=11 \, \Delta q$, in steps of $\Delta q$ where $\Delta q =  \pi/(800 \, d)$.
In (a) the overlap for $q=0$ is identically 0, and the overlap for $q=q_{max}$ has a maximum of approximately 0.66,
whereas in (b) the overlap for $q=0$ is identically 1, and the overlap for $q=q_{max}$ has a minimum of approximately 0.76.
The vertical bars show the bounds $-q_{max}  \le k < 0$, corresponding to the region 
where $| \langle \overline{1}; k^\prime = k+q_{max} | 1; k \rangle | = 1$ in the continuum model.
Outside of this range, $| \langle \overline{1}; k^\prime=k+q_{max} | 1;k \rangle | = 0$ for the continuum model.
The conduction-conduction band has the opposite symmetry in the continuum model.}
\label{fig:overlaps}
\end{figure}

\begin{table}[tp]
\centering
\caption{3nn tight Binding Parameters for the acGNR5 nanoribbon obtained using a fit to EHT \cite{Raza08_ac_prb} data.}
\begin{tabular}{c r l}
\hline
$E_0$ & 0.11031 eV& onsite energy\\
$t_0$ & -2.69341 eV& 1nn hopping parameter\\
$t_1$ & 0.02201 eV& 2nn hopping parameter\\
$t_2$ & -0.03225 eV& 3nn hopping parameter\\
\hline
\end{tabular}
\label{tab:tbparams}
\end{table}

To compute the plasmon dispersion in the random phase approximation (RPA) for the nanoribbon with 3nn Hamiltonian, we follow the procedure outlined in Ref. \cite{dra11a}.  
Due to large energy differences and small electronic wavefunction overlap integrals at $q \approx 0$ for both 1nn pzTB in Ref. \cite{dra11a} and 3nn pzTB in this paper, we use a two-band dielectric function including only the $i=1$ conduction and $i=\overline{1}$ valence bands to study the plasmon dispersion relation. However, some of the details of the 3nn pzTB model are different, which we discuss next.

In the RPA expression for the interband polarizability, electronic wavefunction overlap integrals between states in the two bands at momenta $k$ and $k^\prime = k+q$, where $q$ is the plasmon momentum, play a significant role.  The polarizability is written as,
\begin{align}
\Pi_{mn} (q, \omega) &= \lim_{\eta \rightarrow 0} \frac{g_s}{L_x} \times \nonumber \\  
&\sum\limits_k \frac{f(E_{k m}) - f(E_{k^\prime  n} )}{E_{k m} - E_{k^\prime n } + \hbar \omega + i \hbar \eta}| \langle n; k^{\prime}  | m; k \rangle |^2
\label{eq:chi}
\end{align}
where $m$ and $n$ are band indices, $g_s = 2$ is the spin degeneracy, $L_x$ is the sample length,
$k$ is the momentum of the initial state, $k^\prime = k+q$ is the momentum of the final state,
and $f(E) = 1/[1+e^{(E-\mu)/k_B T}]$ is the Fermi-Dirac distribution function with chemical potential $\mu$ and Boltzmann constant $k_B$, where $T$ is the temperature in $K$. $\hbar$ is the reduced Planck's constant and $\eta$ is a small number. We consider intrinsic acGNRs with the chemical potential $\mu = 0$.
In Fig. \ref{fig:overlaps}, we illustrate several of these overlap integrals as functions of $k$.  It should be noted that the the overlap integral is no longer confined in the region defined by $\mathrm{sign}(k \, k^\prime) = -1$, as it is in the continuum model. Rather, the overlap integral is non-zero well beyond this range.  However, direct transitions at $q=0$ are still forbidden.
The broadening of the overlap integral beyond the hard boundaries of the continuum model 
indicates that there is a coupling to free-carrier states for collective modes at all nonzero $q$, and as a result, plasmons in this system are Landau damped.

Following Ref. \cite{dra11a}, we calculate the longitudinal dielectric function for acGNRs in the RPA. In the RPA, the dielectric matrix for acGNRs can be written as \cite{Zupanovic}:
\begin{equation} 
\epsilon_{ijmn} (q, \omega) = \delta_{im} \delta_{jn} - v_{ijmn} (q) \: \Pi_{mn} (q, \omega)
\label{eq:rpa}
\end{equation}
where $v_{ijmn} (q)$ is the Coulomb matrix element in one dimension, $\Pi_{mn} (q, \omega)$ is the polarizability of the acGNRs, and $i$, $j$, $m$, and $n$ are the band indices. Non-trivial solutions to the field equations require:
\begin{equation}
\det{[\epsilon_{ijmn} (q, \omega)]} = 0
\label{eq:disprel}
\end{equation}

\emph{Intrinsic Plasmons:}
In the two-band approximation for intrinsic acGNRs at $T = 0$, the self-polarizabilities of the $i=\overline{1}, \, 1$ bands are given
as, $\Pi_{\overline{1}\overline{1}} (q, \omega) = \Pi_{11} (q, \omega) = 0$.  Further, symmetries in the acGNRs require\cite{Fertig,Zupanovic} that
the Coulomb matrix elements $v_{\overline{1},1,\overline{1},1} (q) = v_{\overline{1},1,1,\overline{1}} (q) = v_{1,\overline{1},1,\overline{1}} (q) = v_{1,\overline{1},\overline{1},1} (q)$. This result gives the dispersion relation of the collective (plasmon) state in the 2-band approximation by simplifying Eq. \ref{eq:disprel} as follows:
\begin{equation}
1-v_{\overline{1},1,\overline{1},1} (q) \, [ \Pi_{\overline{1} 1} (q,\omega) + \Pi_{1 \overline{1}} (q, \omega) ] = 0
\label{eq:twobanddisp}
\end{equation}
We compute the Coulomb matrix elements $v_{\overline{1}, 1, \overline{1}, 1} (q)$ as described in Ref. \cite{dra11a} using
the $p_z$-orbital wavefunction localization parameter $w=1 \mathrm{\AA}$ \cite{Raza11}.
Solving Eq. \ref{eq:twobanddisp} gives the dispersion relation for the collective modes (plasmons) in the acGNR.

\emph{Extrinsic Plasmons:}
The dispersion relation for plasmons in extrinsic acGNR 
can also be obtained from Eq. \ref{eq:disprel} in the two-band approximation.  For a chemical potential $\mu$ in the $i=1$ conduction band at $T=0$, states with momenta between $-k_f \le k \le k_f$
where $E_{k_f 1} = \mu$ are filled, and states outside of this range are empty.
For the extrinsic case $\Pi_{11} (q, \omega)$ is no longer 0, and we write the plasmon dispersion relation as:
\begin{align}
\left ( 1 - v_{\overline{1},1,\overline{1},1} (q) \, [\Pi_{\overline{1}1} (q, \omega) + \Pi_{1\overline{1}} (q, \omega)] \right ) \nonumber \\
\times \left ( 1 - v_{1,1,1,1} (q) \, \Pi_{11} (q, \omega) \right )= 0
\label{eq:extrinsicdisp}
\end{align}
Plasmons for negative chemical potentials $\mu$ will exhibit similar behavior.

\begin{figure}[htb]
\centerline{\includegraphics[width=8.5cm]{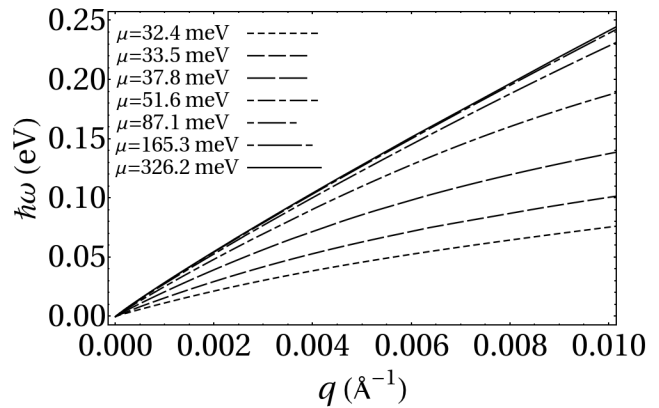}}
\caption{Extrinsic plasmon dispersion relations for acGNR$5$. Dispersion curves are calculated for a range of chemical potentials $\mu$.}
\label{fig:dispsum}
\end{figure}

\section{\label{sec:results}Discussion of Results}

\begin{figure}[htb]
\centerline{\includegraphics[width=8.5cm]{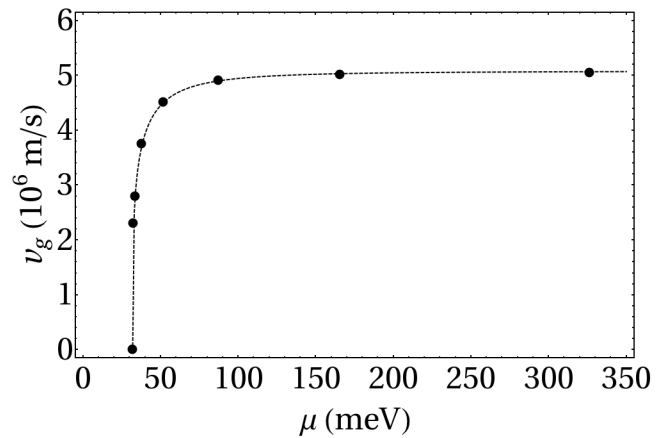}}
\caption{Group velocity of extrinsic plasmons as a function of the chemical potential $\mu$ in the $q \rightarrow 0$ limit.
The solid points are calculated from the dispersion relation data presented in Fig. \ref{fig:dispsum}.  The dashed curve
is calculated using the analytic model for $\lim_{q \rightarrow 0} v_g (\mu)$ discussed in the text
with $v_{1,1,1,1} (0) / (2 e^2 / \epsilon_0) = 11.1294$ and $v_{F1} = 8.33219 \times 10^5$ m/s.}
\label{fig:velocity}
\end{figure}

\emph{Intrinsic Plasmons:}
The intrinsic plasmon obtained using our formalism exhibits an onset threshold in both the $q$ and $E$ dimensions.  The $q$ threshold can be understood with
the data presented in Fig. \ref{fig:overlaps}(a).
For small values of $q$, the overlap integral is nearly zero.  Because the polarizabilities $\Pi_{\overline{1}1} (q, \omega)$ and $\Pi_{1\overline{1}}$ are proportional to this overlap,
the dielectric function never crosses 0, and so no collective mode exists.  As the overlap gets larger, the dielectric function eventually crosses zero and the intrinsic
plasmon dispersion exists.  The threshold in $E$ is a result of the fact that the polarizabilities are not large enough to cause a zero-crossing for small $E$.  As the plasmon
energy increases above the bottom of the conduction band, the resonant enhancement in the polarizabilities causes a zero-crossing.
Because we are interested in plasmons in the $q \rightarrow 0$ limit, we do not consider the intrinsic case further.

\emph{Extrinsic Plasmons:}
Dispersion relations for plasmons in extrinsic acGNR5 computed using the tight-binding formalism described above are also plotted
in Fig. \ref{fig:dispsum} for several values of the chemical potential $\mu > E_g /2$ corresponding to a geometric distribution of $k_F$.  
From these results, it can be readily observed that the dispersion curves have a $q \sqrt{v_{1,1,1,1} (q )}$ character for values of the chemical potential within a few meV
of the band edge ($\mu \gtrsim E_g/2$).
Further, as the chemical potential increases, the dispersion relation is observed to asymptotically approach a limit which corresponds to 
the plasmon dispersion in a massless Dirac fermion system.

It is interesting to analyze the behavior of the extrinsic plasmon group velocity in the $q \rightarrow 0$ limit as a function of the chemical potential $\mu$.
In this limit, the interband polarizabilities $\Pi_{1\overline{1}} (q, \omega)= \Pi_{\overline{1}1} (q, \omega) =0$ because the interband overlap
integral $\langle \overline{1}; k|1; k \rangle = 0$ (see Fig. \ref{fig:overlaps}(a)).
Further, the intraband overlap integral $\langle 1; k | 1; k \rangle = 1$ in this limit (see Fig. \ref{fig:overlaps}(b)).
As a result, the intraband polarizability becomes:
\begin{equation}
\Pi_{11} (q, \omega) = \frac{g_s}{L_x} \sum\limits_k \frac{2 \Delta E}{(\hbar \omega)^2}
\end{equation}
where $\Delta E = E_{k+q,1} - E_{k,1}$.
Thus, as $q \rightarrow 0$, the dielectric function becomes:
\begin{equation}
1-v_{1,1,1,1} (q) \, \Pi_{11} (q, \omega) = 0
\label{eq:simpledielectricfunction}
\end{equation}

Solving Eq. \ref{eq:simpledielectricfunction} for $\omega$, the plasmon group velocity in the $q \rightarrow 0$ limit can then be written:
\begin{equation}
v_g(k_F) = \left [\lim_{q \rightarrow 0} \left (\frac{2 \, v_{1,1,1,1} (q)}{\hbar^2 q^2} \int\limits_{-k_F}^{k_F} \Delta E \, dk \right ) \right ]^{1/2}
\label{eq:velocity}
\end{equation}
where the integral is taken over the filled states between $-k_F$ and $k_F$.
Substituting the relationship between chemical potential and Fermi wavenumber $k_F = \sqrt{\mu^2 - (m_0 v_{F1}^2)^2}/\hbar v_{F1}$ into
the analytic result for Eq. \ref{eq:velocity} gives the group velocity
as a function of chemical potential $v_g (\mu)$,
which is shown as a dashed curve in Fig. \ref{fig:velocity}.

\section{\label{sec:summary}Conclusions}
In summary, we have computed the plasmon dispersion for an acGNR5 nanoribbon using a 3nn tight-binding model.  
This nanoribbon represents a massive Dirac Fermion system.
Hopping parameters for the model were obtained by fitting the 3nn band structure to band data obtained from an EHT
calculation.  The intrinsic plasmon dispersion relation obtained exhibits a threshold in both $q$ and $E$.
The extrinsic plasmon dispersion relation obtained follows the $q \sqrt{V (q )}$ dependence expected in 1D systems for values of the chemical potential near the band edge ($\mu \gtrsim E_g/2$),
and the dispersion relation asymptotically approaches one corresponding to a massless Dirac fermion system as the chemical potential $\mu$ increases.
Good agreement between the group velocity of these plasmons in the $q \rightarrow 0$ limit and an analytic model based on the behavior of the polarizabilities as $q \rightarrow 0$
is obtained.
Finally, we note that some damping of these plasmons may be expected to occur from plasmon scattering to free electron states due to the nature of the
relevant overlap integrals.

\begin{acknowledgments}
DRA acknowledges partial support for this work from the National Institutes of Health. 
\end{acknowledgments}

\bibliography{nano3nn}

\end{document}